\documentclass{basi}
\makeatletter
\def\@maketitle{\newpage
 \vspace*{20pt}
 {\raggedright \sloppy
  {\reset@font\Large\bf \@title \par}
  \vskip 7pt
  {\reset@font\large
    \begin{tabular}[t]{@{}l@{}}\let\\=\author@tabcrone\@author
    \end{tabular}\par}
  \vskip 7pt
 }
 \par\noindent
 {\reset@font\small \@date \par}
 \vskip 6pt
}
\def\author@tabcrone{\vspace{2pt}\tabularnewline[-3pt]\reset@font\small\it
  \let\\=\author@tabcrtwo\ignorespaces}
\def\author@tabcrtwo{\tabularnewline[-3pt]\reset@font\small\it
 \let\\=\author@tabcrtwo\ignorespaces}
\makeatother
\usepackage{graphicx}
\usepackage{txfonts}
\usepackage[longnamesfirst]{natbib}
\usepackage{dcolumn}
\newcolumntype{d}{D{.}{.}{-1}}
\def\dhead#1{\multicolumn{1}{c}{#1}}
\newcommand{\fdg}{.\mskip-1.5\thinmuskip^\circ}
\newcommand{\la}{\,\rlap{\raise 0.5ex\hbox{$<$}}{\lower 1.0ex\hbox{$\sim$}}\,}
\newcommand{\HII}{{\sc H\,ii}}
\pubyear{2007}
\volume{00}
\pagerange{\pageref{firstpage}--\pageref{lastpage}}
\def\todo#1{[[#1]]\strut\vadjust{\kern-\dp\strutbox{\vtop to \dp\strutbox{%
\baselineskip\dp\strutbox\vss\rlap{\hskip\hsize\ \rm{$\Longleftarrow$}}\null}}}
}
\begin{document}

\title[The radio spectrum of HB~3]{Comments on the radio spectrum of HB~3}

\author[D.~A.\ Green]{D.~A.\ Green\\
   Mullard Radio Astronomy Observatory, Cavendish Laboratory,
   19 J.~J.~Thomson Avenue,\\Cambridge CB3 0HE, United Kingdom}

\date{Received 2007 March 22nd}

\maketitle

\label{firstpage}

\begin{abstract}
It has recently been suggested that the radio spectrum of the Galactic
supernova remnant HB~3 shows flattening at higher frequencies (above about
1~GHz). Here I review the radio spectrum of HB~3, noting the difficulties in
deriving accurate flux densities for this remnant, particularly at high
frequencies, due to the proximity of bright, thermal emission from W3 and its
surroundings. A flux density for HB~3 at 2695~MHz is derived from Effelsberg
survey data. The spectrum of HB~3 is well represented by a simple power-law
spectrum from 22 to 2695~MHz, with a spectral index of $0.56 \pm 0.03$. It is
concluded that contamination with thermal emission from adjacent regions is the
cause for the reported spectral flattening of HB~3.
\end{abstract}


\section{Introduction}

Recently \citet{2007ApJ...655L..41U} have proposed -- based largely on the
results compiled by \citet{2005A&A...436..187T} -- that the radio spectrum of
the supernova remnant (SNR) HB~3 ($=$G132.7+1.3) between 22 and 3900~MHz is
curved, showing a flatter spectrum at higher frequencies.
\citeauthor{2007ApJ...655L..41U} suggest that the reported flattening is due to
a thin shell of thermal emission enclosing HB~3, which contributes $\approx
45$~per~cent of the total flux density at 1.4~GHz. Any thin shell contributing
such a large fraction of the flux from the remnant should be apparent from
available observations -- especially those at 1.4~GHz shown in
\citeauthor{2005A&A...436..187T} -- but no such shell is seen. High frequency
flattening of the radio spectra of SNRs has been suggested previously, in the
case of Tycho's and Kepler's SNRs, by \citet{1992ApJ...399L..75R}. In these
cases it was proposed that the spectral flattening was an intrinsic curvature
resulting from the shock acceleration theory, rather than being due to thermal
emission. However, there are several problems with some of the flux densities
of HB~3 used by \citeauthor{2005A&A...436..187T}, particularly that the highest
two frequency observations included (at 3650 and 3900~MHz) have much poorer angular
resolutions than stated, and they are likely to be contaminated with thermal
emission from the nearby bright {\HII} region W3. Here I review the flux
densities available for HB~3, derive a new flux density for it at 2695~MHz from
Effelsberg 100-m survey observations \citep{1990A&AS...85..691F}, and present a
revised radio spectrum for this SNR.

\section{Background}

\subsection{Published flux densities for HB~3}\label{s:published}

\citet{2007ApJ...655L..41U} cite \citet{2005A&A...436..187T} for the flux
densities they use to produce a radio spectrum of HB~3, although they also
include a flux density at 22~MHz, which is not listed by
\citeauthor{2005A&A...436..187T}. This 22-MHz flux density is, apparently,
$450\pm70$~Jy, from \citet[a private communication
from R.~S.~Roger, integrated from the image presented in
\citealt{1969ApJ...155..831R}]{1987AJ.....94..111L}.
There are, however, problems with several of
the flux densities used by \citeauthor{2005A&A...436..187T}, and hence by
\citeauthor{2007ApJ...655L..41U}. First, \citeauthor{2005A&A...436..187T} give
the flux density of HB~3 at 38~MHz as $380\pm80$~Jy, citing
\citet{1967MNRAS.136...11C}. However, the published flux density is actually
$350\pm80$~Jy. Second, \citeauthor{2005A&A...436..187T} use flux densities at
408 and 1420~MHz that they derive from Canadian Galactic Plane Survey (CGPS)
images \citep{2003AJ....125.3145T}, after the removal of the compact
(extragalactic) sources. However, these flux densities are compared with other
flux densities from the literature which have {\sl not} had the compact sources
removed. Therefore, to compare with the other flux densities, it would be
better to use the somewhat higher flux densities at 408 and 1420~MHz that
include the compact sources (i.e.\ 73.9~Jy instead of 68.6~Jy at 408~MHz, and
47.2~Jy instead of 44.8~Jy at 1420~MHz). \citet{2005A&A...436..187T}'s results
contain other errors --  see the errata published in
\citet{2006A&A...451..991T} -- namely the wrong resolution for the Effelsberg
2695-MHz survey observations \citep{1990A&AS...85..691F}, and incorrectly
calculated errors in the spectral index for HB~3 derived from their 408- and
1420-MHz integrated flux densities. These errors also apply to the results for
OA~184 -- which is no longer thought to be a SNR, see
\citet{2006A&A...454..517F} -- and for the SNR VRO 42.05.01 ($=$G166.0$+$4.3),
published in \citet{2005A&A...440..929L}. Another problem, which is difficult
to quantify, is the fact that the flux densities reported in the literature may
not necessarily be on the same flux density scale. Most recent observations, at
the higher frequencies, are likely to be on the scale of
\citet{1977A&A....61...99B}. However, this scale does not provide secondary
calibrator flux densities at frequencies below 400~MHz, and there is evidence
that the secular change in spectral index used by
\citeauthor{1977A&A....61...99B} for the absolute calibrator Cas~A is incorrect
-- see, for example, \citet{1990MNRAS.243..637R} -- which is relatively more of
a problem for the low frequency flux densities available for HB~3 in the
literature. However, the main problem with making a direct comparison of the
published flux densities of HB~3 is due to the fact that this SNR is close to
the bright thermal complex W3/4/5, which is discussed in more detail below.

\begin{figure}
\centerline{\includegraphics[height=7cm]{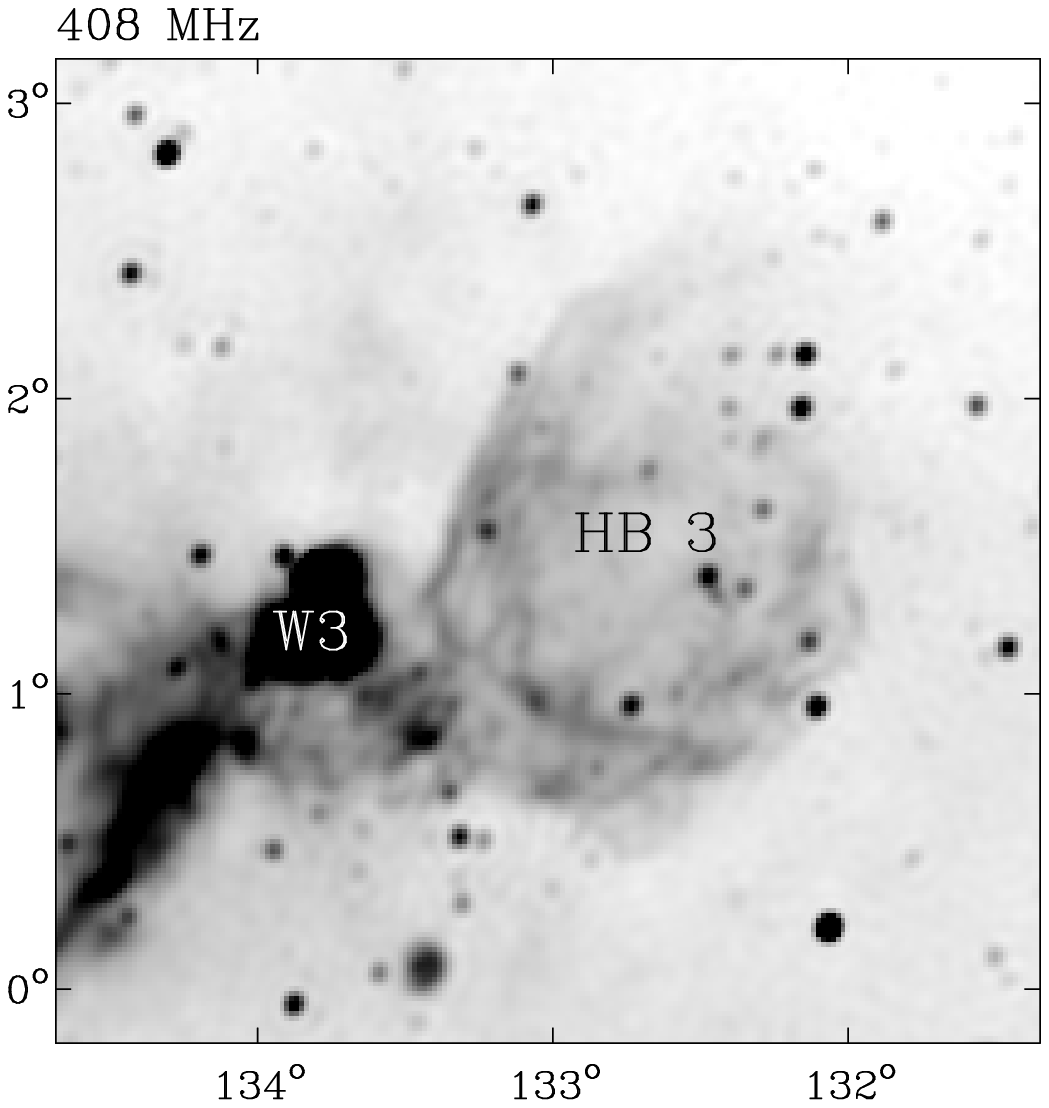}\quad
            \includegraphics[height=7cm]{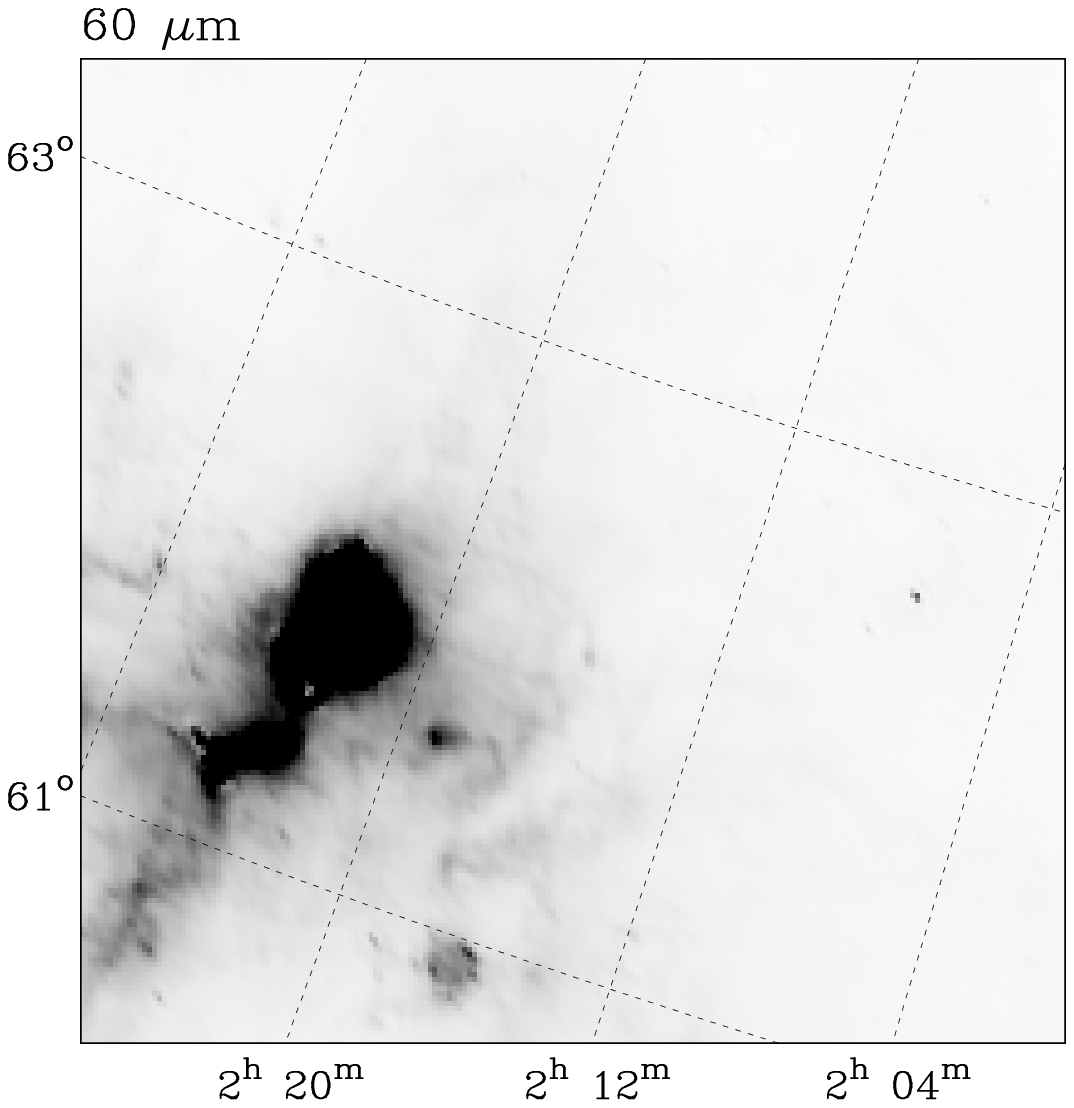}}
\caption{(left) CGPS image at 408~MHz of the region around HB~3, labelled with
Galactic coordinates. The resolution is $3.4 \times 3.8$ arcmin$^2$ (${\rm EW
\times NS}$), and the greyscale of brightness temperature is from 50 to 150~K
(i.e.\ 0.32 to 0.96 Jy~beam$^{-1}$. (right) IRAS image at 60~$\mu$m of same
region, labelled with RA and Dec (B1950.0). The greyscale is from 0 to 300
MJy~sr$^{-1}$.\label{f:74cm60micron}}
\end{figure}

\subsection{The extent of HB~3}\label{s:extent}

Deriving accurate flux densities for HB~3 is difficult because it is adjacent
to the {\HII} regions W3/4/5 (e.g.\ \citealt{1980ApJ...238..829D,
1991A&AS...87..177R}; \citealt{1997ApJS..108..279N}). W3 ($=$IC 1795) is a very bright
{\HII} region -- near $l=133\fdg8$, $b=1\fdg2$, to the east of HB~3.
Fig.~\ref{f:74cm60micron} shows images of HB~3 and its vicinity at both a
frequency of 408~MHz and a wavelength of 60~$\mu$m. The 408-MHz image is from
the CGPS, and is the same data as used by \citeauthor{2005A&A...436..187T}.
This image shows the extent of the remnant better than do higher radio
frequency observations (see Fig.~\ref{f:11cm} below). The 60-$\mu$m image is
from IRAS observations, reprocessed for comparison with CGPS radio observations
(see \citealt{1997ApJS..111..387C}), and shows the extent of thermal emission
in the region. The resolution varies across the 60-$\mu$m image, but is
typically $\sim 1.6 \times 1$ arcmin$^2$ (PA $\sim 35^\circ$). The confusion of
HB~3 with W3 is particularly an issue for lower resolution observations, where
separating out the remnant from W3 is difficult, and more so at higher
frequencies when the non-thermal emission from the remnant is relatively
fainter (see Fig.~\ref{f:11cm} below).

Indeed, the difficulty of defining the extent of HB~3 in the presence of the
nearby thermal emission has been noted by various authors.
\citet{1987AJ.....94..111L} comment that there is ``no definite boundary
separating it [HB~3] from W3'', and that ``There is a plateau of emission to
the south and west of W3'', which they go on to show is probably mostly
thermal. \citet{2003A&A...408..961R} present observations of HB~3 at 865~MHz,
with a resolution of $14.6 \times 14.3$ arcmin$^2$, together with observations
of many other SNRs at this frequency. In the case of HB~3 they note that it is
difficult to obtain an accurate integrated flux density for the remnant due to
the complex surroundings of the remnant\footnote{Although they themselves give a
relatively accurate flux density for HB~3 at 865~MHz.}, and go on to
say ``This limitation holds not only for the present observations, but also for
most of the previously published images with larger beam sizes.'' This
difficulty of separating out the emission from HB~3 from the thermal emission
associated with W3 is particularly a problem at higher frequencies. The
reported spectral flattening of HB~3 depends almost solely on high frequency
observations of \citet{1987AISAO..25...84T}. However,
\citeauthor{2005A&A...436..187T} erroneously give the resolution of the 3650-
and 3900-MHz observations of \citeauthor{1987AISAO..25...84T} as $1 \times 10$
arcmin$^2$ at each frequency. In fact the resolution of these observations is
considerably larger: $1.3 \times 46$ arcmin$^2$ at 3650~MHz and $1.2 \times 43$
arcmin$^2$ at 3900~MHz (${\rm EW \times NS})$. Given that the RATAN-600 telescope used
for these observations makes scans at fixed declinations, the large
north--south extent of the beam means that contamination with thermal emission
for the eastern part of the remnant -- i.e.\ from the plateau region south and
west of W3, as described by \citeauthor{1987AJ.....94..111L} -- seems
unavoidable.

\citeauthor{1987AISAO..25...84T} also present observations of HB~3 at 960~MHz,
which have a large beam size, of $4.7 \times 165$ arcmin$^2$. Unsurprisingly,
contamination with nearby thermal emission is a serious problem with the
960-MHz flux density for HB~3 from these observations. The reported flux
density of $70\pm5$~Jy is high by a factor of about 1.45 compared with what
might be expected given other available flux densities at lower and higher
frequencies (cf.\ Fig.~\ref{f:spectrum} below). Moreover, it is not clear that
the observations of \citeauthor{1987AISAO..25...84T} were made with a
sufficient number of scans to properly sample the emission from HB~3 in
declination. The contamination of the 960-MHz flux density from
\citeauthor{1987AISAO..25...84T} was noted by \citeauthor{2005A&A...436..187T},
who did not use this flux density for their spectrum. They also noted that
the 3650- and 3900-MHz observations may be contaminated with thermal emission.
Rather than discard the 3650- and 3900-MHz flux densities -- as they did with
the 960-MHz flux density value -- \citeauthor{2005A&A...436..187T} used the
published flux densities, but increased their errors to 30~per~cent. Since any
thermal contamination is a {\em systematic} increase in the flux density, it is
inappropriate to simply increase the ({\em random}) errors of these flux
densities.

Discarding the RATAN-600 observations, the highest available frequency
with a measured flux
density is 2695~MHz, from the Effelsberg survey
(\citeauthor{1990A&AS...85..691F}). Given the potential problems of
contamination with nearby thermal emission, it is important to define the
extent of the remnant well. From Fig.~\ref{f:74cm60micron} it is clear that it
is not easy to separate the emission from HB~3 from the emission from W3 and
its surroundings -- particularly to the south and west of W3 -- as has been
noted by previous authors. Indeed, \citeauthor{1987AISAO..25...84T} themselves
note that their baselevels are less reliable in the eastern part of the
remnant. They give separate flux densities for the whole remnant, from $02^{\rm h}
08^{\rm m}$ to $02^{\rm h} 20^{\rm m}$ (B1950), and for the western part
alone, $02^{\rm h} 08^{\rm m}$ to $02^{\rm h} 16^{\rm m}$ only. From their
observations they report a steeper spectrum for the western part of the remnant
than for the remnant as a whole, which is consistent with thermal
contamination from W3 and its surroundings in the east. Thus, the flux densities
reported by \citeauthor{1987AISAO..25...84T} for HB~3 as a whole do indeed
appear to be significantly contaminated by thermal emission in the east. Note,
from the IRAS image shown in Fig.~\ref{f:74cm60micron}, that there is emission to
the south and west of W3. This is presumably thermal emission near $l \approx
133\fdg5$, $b\approx 0\fdg9$, which is west of $02^{\rm h} 20^{\rm m}$.

\begin{figure}
\centerline{\includegraphics[height=7cm]{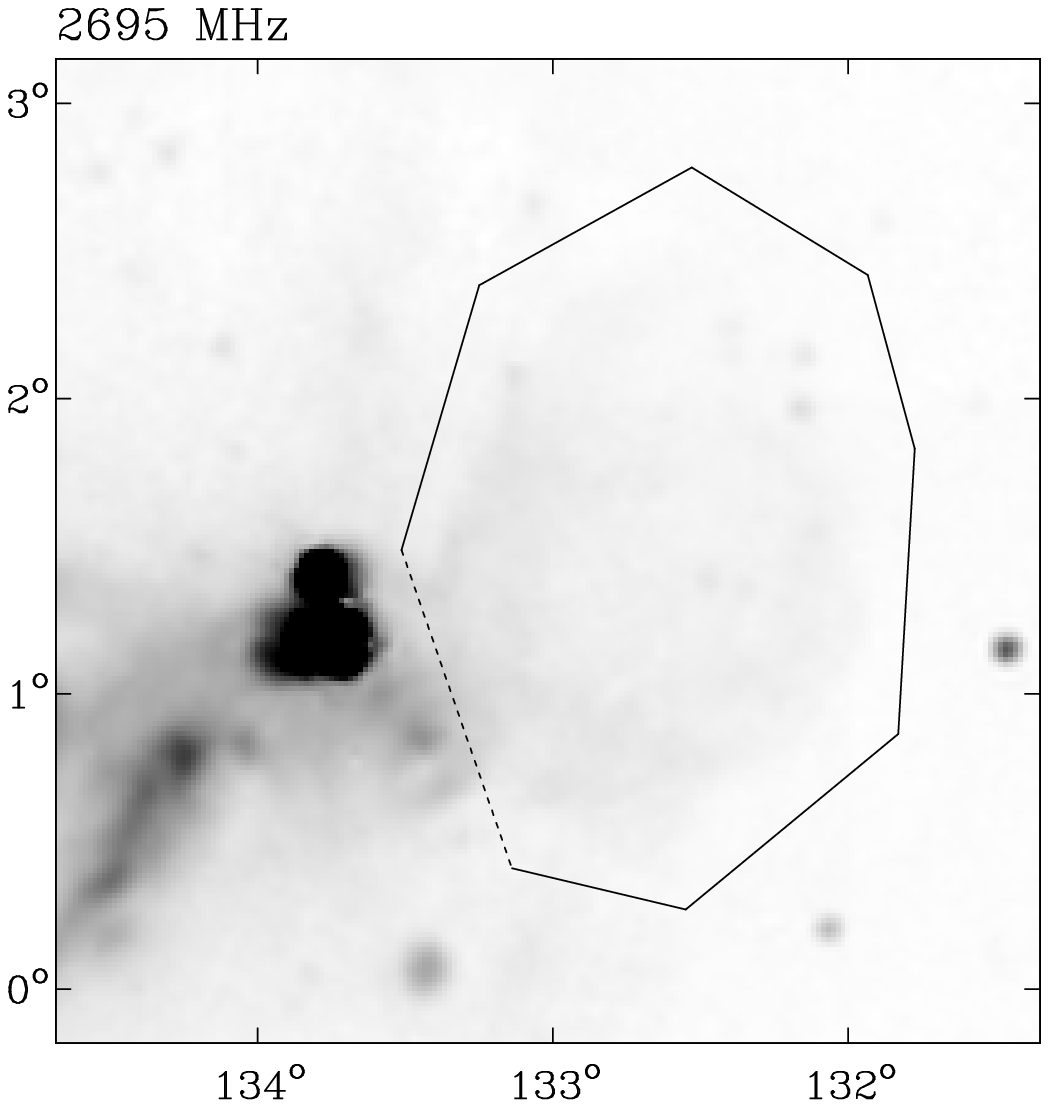} \quad
            \includegraphics[height=7cm]{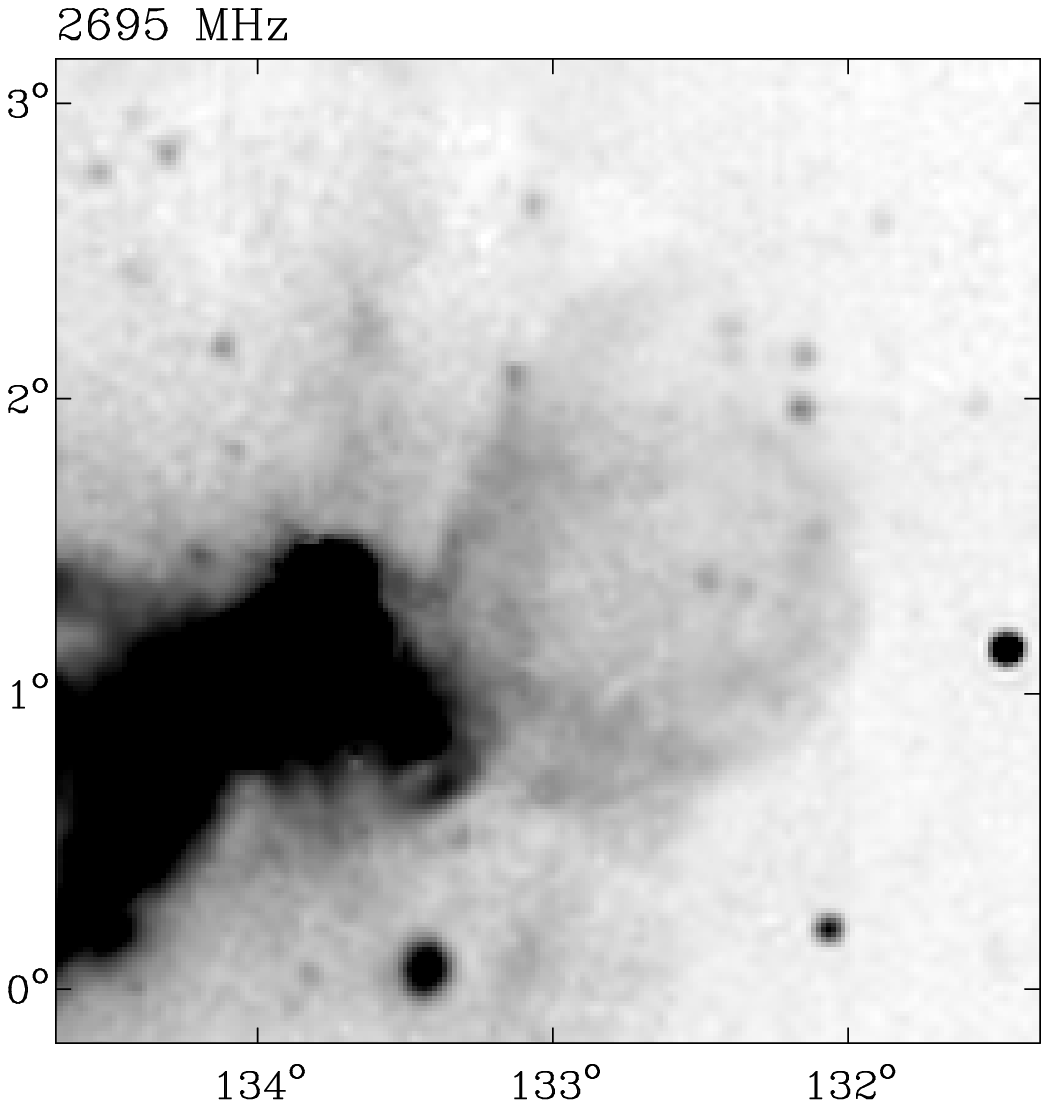}}
\caption{Effelsberg 2695-MHz image of HB~3, with a resolution of $4.3$~arcmin:
(left) showing the brighter emission, with a greyscale range 0 to 6~K (i.e.\ 0
to 2.4 Jy~beam$^{-1}$), (right) showing the fainter emission, with a greyscale
range of 0 to 1.5~K (i.e.\ 0 to 0.59 Jy~beam$^{-1}$). The peak brightness in
the image,
from W3, is 81~K. The polygon shows the region for which the integrated flux
density for HB~3 was derived, after removal of a twisted plane fitted to the
solid edges of the polygon (see Section~\ref{s:new}).\label{f:11cm}}
\end{figure}

\section{A 2695-MHz flux density for HB~3}\label{s:new}

Given the problems with the RATAN-600 data -- and the misquoted resolution
given by \citeauthor{2005A&A...436..187T} for the Effelsberg 2695-MHz survey
data -- I have derived an integrated flux density for HB~3 from this survey.
HB~3 and its surroundings at 2695~MHz are shown in Fig.~\ref{f:11cm}, with
different greyscales, to show both the bright emission from W3, and the much
fainter emission from HB~3. To derive an integrated flux density for HB~3, the
extent of the remnant was defined from the CGPS 408-MHz image
(Fig.~\ref{f:74cm60micron}), by drawing a polygon around the SNR, as is
reproduced in Fig.~\ref{f:11cm}. It should be recognised, however, that the
dividing line between the remnant and the thermal emission in the south and
west of W3 is subjective. In order to derive a flux density for the emission
within the polygon, a twisted plane was first fitted to the pixels lying around
the edges of the polygon, omitting the line that divides the remnant from the
extended emission in the south and west of W3. This twisted plane was then
removed from the image before integrating the emission within the polygon. The
integrated flux density of HB~3 at 2695~MHz, inside the polygon shown on
Fig.~\ref{f:11cm}, is 31~Jy. The integrated flux density varies by about 1~Jy
(i.e.\ about 3~per~cent) if slightly different polygons are chosen, with
similar dividing lines between HB~3 and W3. However, to be cautious, and to
reflect any uncertainties in the overall flux density scale of the Effelsberg
2695-MHz survey data, I will assume an overall uncertainty of 15~per~cent,
giving a 2695-MHz flux density of $31.0 \pm 4.6$~Jy for HB~3. This value is
slightly lower than other flux densities available at this frequency, but not
significantly so (cf.\ the value of $34.7 \pm 12.9$~Jy derived by
\citeauthor{2005A&A...436..187T} from the same survey data, and the value of
$33.8 \pm 7.0$~Jy from \citealt{1974A&A....32..375V}).

If the dividing line between HB~3 and W3 -- i.e.\ the dashed line shown on
Fig.~\ref{f:11cm} -- is changed to include the emission south and west of W3
that is east of $02^{\rm h} 20^{\rm m}$, the derived integrated flux density
increases considerably, to about 40~Jy. This implies that the 3650- and
3900-MHz flux densities reported by \citeauthor{1987AISAO..25...84T} are indeed
likely to be too high, as discussed in Section~\ref{s:extent}.

\begin{table}
\caption{Radio flux densities for HB~3. See discussion in
Section~\ref{s:discuss}.}\label{t:fluxes}
\def\noflux{\multicolumn{2}{c}{}}
\tabcolsep2pt\smallskip\centering
\begin{tabular}{dd@{\kern-6pt$\times$\kern4pt}dd@{\kern-6pt$\pm$\kern4pt}ddd@{\kern-12pt$\pm$\kern4pt}dl}\hline
\dhead{}      & \dhead{}   & \dhead{}            & \multicolumn{2}{c}{published}    & \dhead{flux}    & \multicolumn{2}{c}{scaled}       &           \\
\dhead{$\nu$} & \multicolumn{2}{c}{beamsize}     & \multicolumn{2}{c}{flux density} & \dhead{scaling} & \multicolumn{2}{c}{flux density} & reference \\
\dhead{/ MHz} & \multicolumn{2}{c}{/ arcmin$^2$} & \multicolumn{2}{c}{/ Jy}         & \dhead{factor}  & \multicolumn{2}{c}{/ Jy}         &           \\ \hline
  22 & 102   & 66   & 450   & 70   & 1.05 & 472 & 74 & notes (a) and (b)              \\
  38 &  45   & 45   & 350   & 80   & 1.03 & 360 & 82 & Caswell (1967), note (a)       \\
  83 &  59   & 31   & 180   & 30   &      & \noflux  & Kovalenko et al.\ (1994)       \\
 102 &  48   & 25   & 165   & 30   &      & \noflux  & Kovalenko et al.\ (1994)       \\
 111 &  44   & 23   & 155   & 30   &      & \noflux  & Kovalenko et al.\ (1994)       \\
 178 &  23   & 19   & 120   & 20   & 1.11 & 133 & 22 & Caswell (1967), note (a)       \\
 232 &   3.8 &  4.3 & 120   & 20   &      & \noflux  & note (c)                       \\
 408 &   3.4 &  3.8 &  73.9 & 11.9 &      & \noflux  & Tian \& Leahy (2005); note (d) \\
 610 &  16   & 20   &  60   & 18   &      & \noflux  & note (e)                       \\
 865 &  14.6 & 14.3 &  51.5 &  7.7 &      & \noflux  & Reich et al.\ (2003); note (f) \\
1420 &   1   &  1.1 &  47.2 & 12.2 &      & \noflux  & Tian \& Leahy (2005); note (d) \\
2695 &   4.3 &  4.3 &  31.0 &  4.6 &      & \noflux  & this paper                     \\ \hline
\end{tabular}
\medskip
\begin{minipage}{13cm}
\small Notes: (a) the published flux density has been rescaled to be on the
scale of Baars et al.\ (1977); (b) private communication from R.~S.~Roger to
Landecker et al.\ (1987); (c) from MSci thesis of J.~Zhu (1993), reported in
Tian \& Leahy (2005); (d) including compact sources; (e) private communication
from H.~Wendker to Landecker et al.\ (1987); (f) the error has been increased
to a nominal 15~per~cent.
\end{minipage}
\end{table}

\begin{figure}
\centerline{\includegraphics[width=13.5cm]{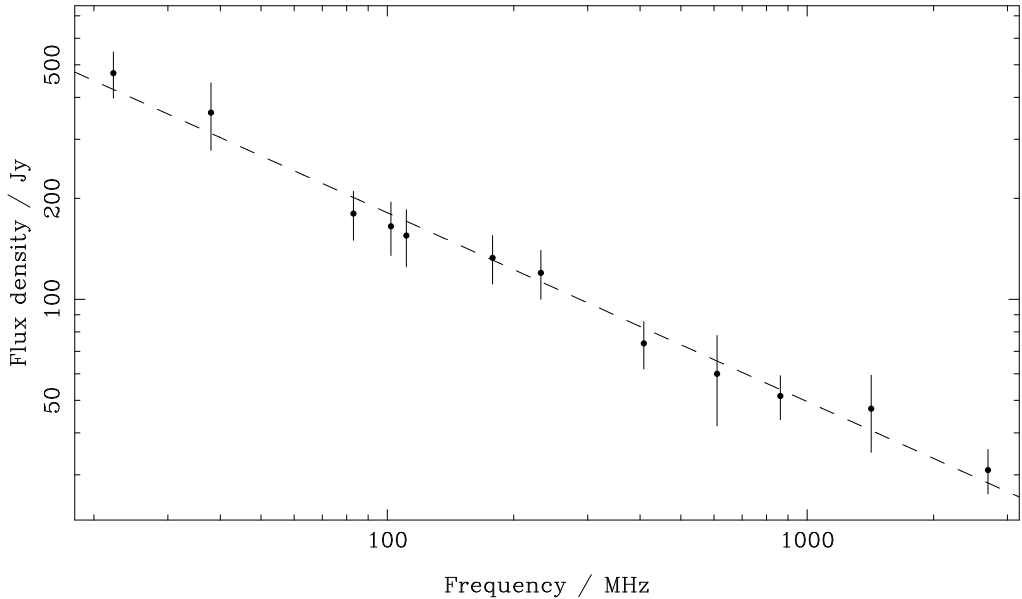}}
\caption{A revised radio spectrum for HB~3, from the flux densities given in
Table~\ref{t:fluxes}. See Section~\ref{s:discuss} for some notes on the flux
densities used or not used, and which have been rescaled slightly. The dashed
line is the best fitting power law, with $S \propto
\nu^{-0.56}$.}\label{f:spectrum}
\end{figure}

\section{Discussion and Conclusions}\label{s:discuss}

Table~\ref{t:fluxes} gives integrated flux densities for HB~3. These are as
listed by \citeauthor{2005A&A...436..187T}, but with the following revisions:
(i) a 22-MHz flux density, from a private communication from R.~S. Roger to
\citet{1987AJ.....94..111L}; (ii) the 38-MHz flux density has been corrected to
the published value (see Section~\ref{s:published}); (iii) at 408 and 1420~MHz,
the values for the flux density of HB~3 from \citeauthor{2005A&A...436..187T}
before removal of the compact sources are preferred (again see
Section~\ref{s:published}); (iv) the relatively small uncertainty at 865~MHz
quoted by \citet{2003A&A...408..961R} has been increased to a nominal
15~per~cent, with the other uncertainties as quoted in the literature, which
are taken to be $1\sigma$ errors; (v) the 2695-MHz flux density derived in
Section~\ref{s:new} is preferred to other flux densities available at similar
frequencies; (vi) the 3650- and 3900-MHz observations from
\citeauthor{1987AISAO..25...84T} are omitted due to their likely contamination
with thermal emission (see discussion in Sections~\ref{s:extent} and
\ref{s:new}); (v) the older flux densities have been scaled to place them in
the scale of \citeauthor{1977A&A....61...99B}\footnote{The 22-MHz flux density
is, presumably, on the scale used by \citet{1986A&AS...65..485R}, which is
based on an assumed flux density of 29,100~Jy for Cygnus~A. This is 5~per~cent
below the value given by the absolute spectrum for Cygnus~A from
\citeauthor{1977A&A....61...99B} The 38-MHz flux density reported by
\citeauthor{1967MNRAS.136...11C} is from the survey data of
\citet{1967MmRAS..70...53W}, which is on the scale of
\citet{1963MNRAS.125..261C}, as is the 178-MHz flux density. The correction
factors given by \citeauthor{1977A&A....61...99B} for the CKL scale have been
used for the 38- and 178-MHz flux densities.}, although -- as noted in
Section~\ref{s:published} -- there are problems with this scale at low
frequencies. The flux density scales used for the 232- and 610-MHz values are
not clear, but these are likely to be on the scale of
\citeauthor{1977A&A....61...99B} Note also that all the available flux
densities at 178~MHz or below have been made with poor resolution, so that
obtaining flux densities from these observations is difficult, because of
confusion with nearby thermal emission -- see Section~\ref{s:extent} above,
although this is likely to be less of a problem than at gigahertz frequencies
-- or, at 22 and 38~MHz, because the thermal emission is absorbed (see
\citealt{1967MNRAS.136...11C,1969ApJ...155..831R}).
HB~3 and its surroundings are also imaged at 151-MHz in part of the 6C survey
\citep{1993MNRAS.262.1057H}, with a resolution of $\sim 4$~arcmin, and by
\citet{1998MNRAS.294..607V}, with a resolution of $\sim 1$~arcmin. However,
neither of these surveys image extended sources well, so have not been used to
determine a flux density for HB~3.

Figure~\ref{f:spectrum} shows the radio spectrum from the flux densities given
in Table~\ref{t:fluxes}, with the least squares single power law fit to the
observations. These results show that a single power law fits the data well,
with no obvious flattening of the radio spectrum of HB~3 at high frequencies.
Since the best fit straight line goes through {\em all} the error bars, this
suggests that some of the errors have been overestimated. The least squares fit
gives a spectral index $\alpha$ -- here defined in the sense that flux density
$S$ scales with frequency $\nu$ as $S \propto \nu^{-\alpha}$ -- of $0.56 \pm
0.03$. From the discussion above, the flattening reported by
\citet{2005A&A...436..187T}, and used by \citet{2007ApJ...655L..41U}, seems to
be due to contamination of the higher frequency RATAN-600 flux densities by
thermal emission associated with W3, not thermal emission associated with the
remnant.

It is generally difficult to derive accurate radio spectra for most Galactic
SNRs, due to the potential confusion with other Galactic sources, and also
variations in the Galactic background. Ideally, reasonable resolution
observations are needed over a wide range of frequencies, but in practice -- as
seen here in the case of HB~3 -- these are generally not available at low
frequencies (below a few hundred megahertz), nor at high frequencies (above a
few gigahertz).

\section*{Acknowledgements}

The research presented in this paper has used data from the Canadian Galactic
Plane Survey, a Canadian project with international partners, supported by the
Natural Sciences and Engineering Research Council. I am also grateful to all
those who have been involved with writing the DRAO `export software' package,
particularly the {\sc madr} and {\sc plot} programs, which have been used for
this work. I thank Stanislav Shabala for his translation from the Russian of
the paper by Trushkin et al., and the referee for useful comments on the
earlier version of this paper.

\setlength{\bibsep}{0pt}
\def\newblock{}

\label{lastpage}
\end{document}